\documentclass[draft]{agujournal2019}
\usepackage{url} 
\usepackage{soul}
\usepackage{subfig}
\usepackage{upgreek}
\draftfalse

\journalname{Geophysical Research Letters}

\begin{document}

%
%

\title{The temporal limits of predicting fault failure}


%
%




\authors{Kun Wang\affil{1,2}\thanks{NOTE: KW and CWJ contributed equally to this paper},
         Christopher W. Johnson\affil{1*},
         Kane C. Bennett\affil{1},
         Paul A. Johnson\affil{1}}

\affiliation{1}{Los Alamos National Laboratory, Geophysics Group, Los Alamos, N.M.}
\affiliation{2}{Center for Nonlinear Studies, Los Alamos National Laboratory, Los Alamos, NM, 87545, USA}






\correspondingauthor{Paul A. Johnson}{paj@lanl.gov}




\begin{keypoints}
\item The limits of machine learning models are explored for predicting the future frictional behavior of laboratory earthquake slip
\item A deep learning model is applied to acoustic emission data broadcast from the fault to predict future friction
\item Acoustic emissions broadcast from the fault are found to carry information used for predicting future fault friction
\end{keypoints}

%
%

%
%


\begin{abstract}

Machine learning models using seismic emissions can predict instantaneous fault characteristics such as displacement in laboratory experiments and slow slip in Earth. Here, we address whether the acoustic emission (AE) from laboratory experiments contains information about near-future frictional behavior. The approach uses a convolutional encoder-decoder containing a transformer layer. We use as input progressively larger AE input time windows and progressively larger output friction time windows. The attention map from the transformer is used to interpret which regions of the AE contain hidden information corresponding to future frictional behavior. We find that very near-term predictive information is indeed contained in the AE signal, but farther into the future the predictions are progressively worse. Notably, information for predicting near future frictional failure and recovery are found to be contained in the AE signal. This first effort predicting future fault frictional behavior with machine learning will guide efforts for applications in Earth.
\end{abstract}

\section*{Plain Language Summary}
Over the last 5 years it has been shown that machine learning is a powerful tool to learn about the inside of faults from the acoustic emissions that the fault broadcasts. For instance, the emissions can inform us of the instantaneous fault displacement and friction in addition to timing of an upcoming earthquake in laboratory experiments as well as for the phenomenon of slow slip in Earth. Here we show that by applying a deep learning approach that has shown striking results in natural language processing and computer vision, the seismic emissions also contain foreshadowing information about the immediate future of fault friction. Remarkably, the emissions can tell us if the laboratory fault is about to fail, and how it will begin to recover.

%
%

%


%
%
%
%

\section*{Introduction}
Applying machine learning techniques to fault slip from laboratory shear experiments has demonstrated that continuous acoustic/seismic signals emanating from an active fault contain rich information regarding the instantaneous and future characteristics of that system. The fault acoustic emissions are imprinted with information regarding its current state and where it is in the slip cycle. Indeed, the statistical features of the continuous seismic signal emitted from the fault and identified by the machine learning model allow the prediction of instantaneous fault friction, displacement, fault gouge thickness, slip velocity and the timing of the next laboratory earthquake ('labquake'), e.g., \cite{Rouet-Leduc_GRL_2017,Rouet-Leduc_GRL_2018,Lubbers_GRL_2018,Hulbert_NatGeo_2019,Hulbert_NatGeo_2019,JohnsonPA_PNAS_2021,Wang2021}. The labquake magnitude can also be predicted to a lesser degree, and with considerably less precision and accuracy than quantities related to timing \cite{Hulbert_NatGeo_2019}.   

Other model inputs such as seismic wave velocity, wave amplitude \cite{Shokouhi2021}, and geodetic measurements \cite{Corbi2019} have also been successfully applied to predict instantaneous and future fault characteristics. Since the seminal paper first describing machine learning model predictions by Rouet-LeDuc and colleagues \cite{Rouet-Leduc_GRL_2017}, many works now exist describing these behaviors---too many to reference here. In addition, a Kaggle competition based on laboratory earthquake prediction  was also recently held where many hundreds of competitors took part in developing machine learning models for predicting timing of the upcoming slip event \cite{JohnsonPA_PNAS_2021}.

In previous works, only predictions of the instantaneous characteristics and/or timing (magnitude) of the next slip event were attempted. The question we address here is the following---is there information contained in the continuous seismic signal regarding the near future fault behavior including and beyond the next slip event? One might expect for instance, that due to a slip event the granular gouge is reset due to the intense, system wide perturbation associated with the labquake.  We have observed previously that intense, externally applied dynamic perturbations--dynamic earthquake triggering--can have persistent impact through multiple slip cycles \cite{Johnson2008}. Such perturbations are considerably stronger than the seismic emission associated with the slip event however. Therefore it is unclear whether or not fault slip physical information can be carried within the fault gouge through a slip event into the next slip cycle. 

Predicting future physical behavior in the form of bulk friction from current seismic emissions  is the focus of this work.  Such work will ultimately guide efforts in Earth to determine if there is near-term predictive information regarding fault physical characteristics such as displacement and earthquake timing contained in the continuous  seismic emissions from active faults.

In the following, we describe the laboratory experiment and the accompanying data, and the deep learning modeling applied to those data. We then describe the prediction results and close with discussion of the results and conclusions.

\section*{Methods}

\subsection*{Deep Learning Model}
The deep learning model developed here for the future activity in the laboratory fault experiments is based on the Transformer architecture \cite{vaswani2017attention} designed as an encoder-decoder, or sequence-to-sequence, model that has shown remarkable results in natural language processing, computer vision, and time series modeling applications \cite{dosovitskiy2020image, li2019enhancing, zhou2021informer}. 
This particular model type was selected because the self-attention mechanism greatly improves the performance on sequential data compared to earlier designs that combine an attention model with recurrent neural networks such as Long Short-Term Memory or Gated Recurrent Units \cite{bahdanau2014neural, luong2015effective}. 
The success of the future prediction task relies on the ability of the Transformer model to learn a long-range dependency. 
Prior applications using Transformer models for long-sequence time-series forecasting have shown good results up to about 1000 future time points. 
These improvements rely on more efficient self-attention mechanisms such as LogSparse \cite{li2019enhancing} and ProbSparse \cite{zhou2021informer} that reduce computational complexity and memory usage. 

In this work, the deep learning model is designed to predict the future friction coefficient as model output using as input acoustic emission (AE) obtained from laboratory biaxial shear data. 
Our approach uses a simplified version of the Transformer model that takes a time series vector as input and encodes it to a high dimensional representation of the signal that is then fed to the latent space tensor. 
The coefficients are passed through the Transformer and output to a second latent space. A decoder branch outputs friction coefficients at the corresponding future time step. 

\begin{figure}[htbp]
\centering
\includegraphics[width=0.75\linewidth]{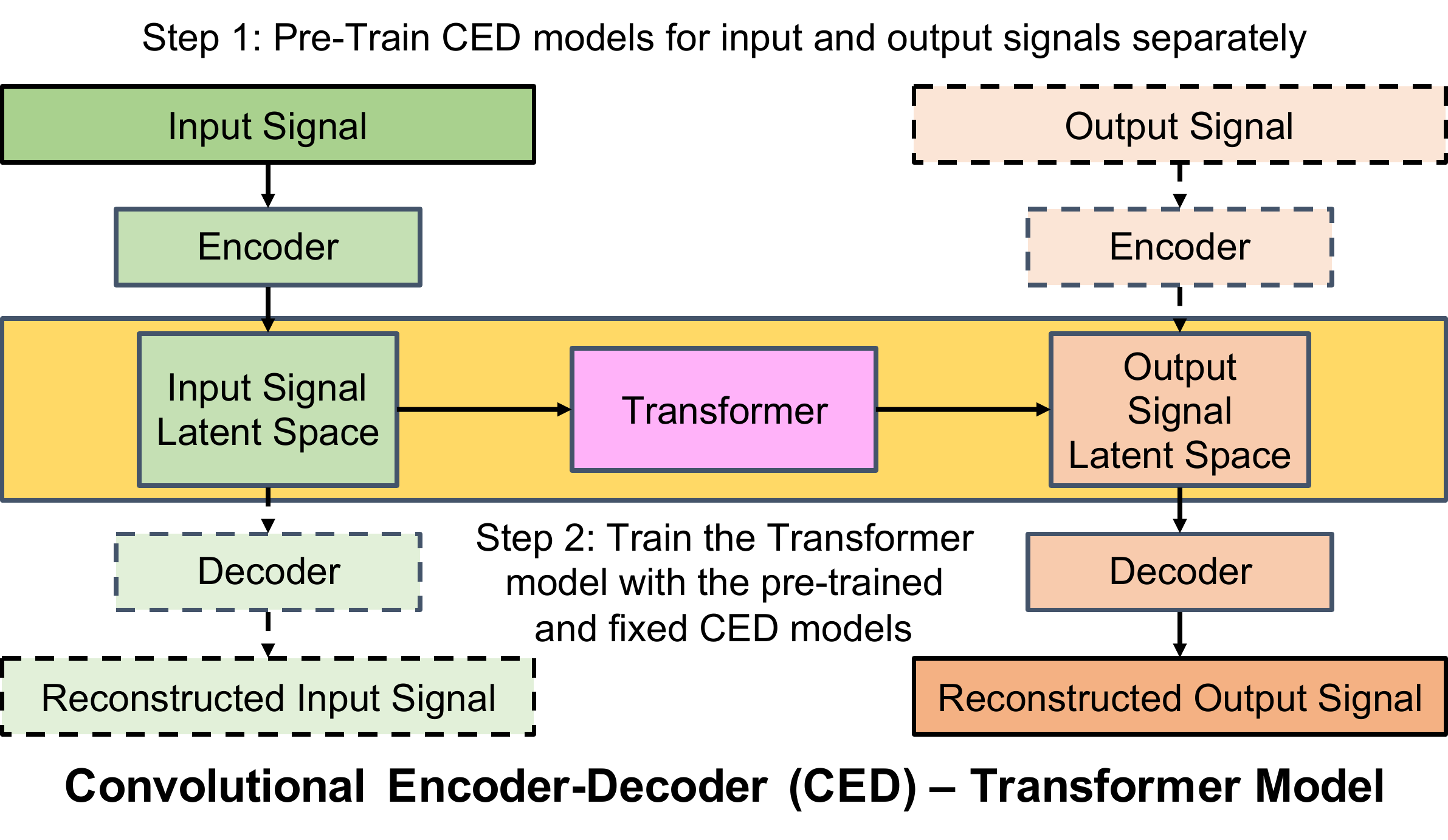}
\caption{The deep learning model architecture used for pre-training and application to future predictions. The input signal (green column) and output signal (pink column) show the convolutional encoder-decoders that are connected by the Transformer via the latent space. The encoder consists of convolutional layers that take the input acoustic emission time series and map it into a sequence of vectors in a smaller space. The vectors are fed into a Transformer (yellow box) to predict output latent vectors and then to a decoder which transforms them into an output sequence with the target fault friction. The model components noted by the dashed lines are used in Step 1 for pre-training as an autoencoder to obtain the embedding vectors and not used in Step 2 with the Transformer for future predictions.}
\label{fig:ced_model_architechture}
\end{figure}

The first step is to design the input and output branches of the model as a convolutional encoder-decoder (CED) model and independently pre-train as an autoencoder (Figure \ref{fig:ced_model_architechture}). 
The input-branch decoder is needed to obtain the embedding of the unit window of signals for a functional latent space. 
The output-branch encoder is needed for generating the target $\mu$ vectors for the teacher-forcing training stage of the Transformer. 
Note that the input decoder and the output encoder (the blocks in dashed lines in Figure \ref{fig:ced_model_architechture}) are not used in the Transformer model application, but only for model pre-training. 
Following the pre-training procedure, the parameters of the input and output CED models are fixed and the Transformer alone is trained using the future friction prediction data.

The latent vectors of the input AE signals are obtained by the pre-trained encoder model branch. 
For example, a total AE history of 2.56 s is encoded into a sequence of 10 input latent vectors, each corresponding to 0.256 s windows of the input signal, with an embedding dimension of 64. 
The self-attention is the learned score representing the time-dependent relative importance of the input signal to the model for each prediction. 
The latent vectors are computed using a multi-head attention layer \cite{vaswani2017attention} in the Transformer's encoder. 
These self-attention processed AE vectors are the keys and values to the multi-head attention in the Transformer's decoder and the embedding vectors of the output $\mu$ generated in the past are the queries.  
The Transformer decodes to the next output embedding vector to match the target $\mu$ vector. 

 The approach applied here uses categorical predictions that represent different embedding, or latent, vectors from the groups of 256 signal data. 
 This is analogous to translating input texts to output token IDs for natural language processing with a Transformer model except that here the dictionary terms are replaced with groups of 256 signal output data points. 
 The alternative approach, which was tested during model development, is to predict embedding vectors of dimension 64 for a regression analysis. 
 Hereafter, we refer to using the  model to predict latent vectors of output friction coefficients as the regression approach and predicting category IDs of the outputs as the classification approach. 
 The advantage of the classification approach is that the Transformer only needs to predict the probability logits, with the maximum value corresponding to the highest probability category. 
 In the regression approach, alternatively, the Transformer needs to predict all dimensions of the latent vectors accurately to successfully estimate the output values. 
 We obtain categories of the output embedding vectors through the Vector Quantised Variational AutoEncoder (VQ-VAE) \cite{oord2017neural} that learns a codebook of discrete vectors in the latent space.
 All continuous vectors after the convolutional encoder are compared against the codebook and the discrete latent vector with the smallest euclidean distance is fed to the convolutional decoder. 
 The codebook is updated by an exponential moving average \cite{oord2017neural} during the training stage. 
 In a classification model the target categories are the discrete latent vectors of the codebook.

The model training is performed in 2 steps. 
In Step 1 the input and output CED branches are trained independently using the mean squared error (MSE) loss function between the target and reconstructed signal, plus the loss associated to the commitment loss in the VQ-VAE and the L2 regularization. 
In Step 2 the Transformer is trained using the classification approach for imbalanced data which focuses more on the categories that are difficult for the model to predict \cite{lin2017focal}, since the number of values associated with the friction drops is much smaller than when the friction increases or is stable. 
Additionally, Step 2 is tested by training the Transformer using the regression approach with the MSE loss function between the target latent vectors of the output friction and the predicted vectors. 
In both training steps a batch size of 8 is used with the Adam optimizer and a custom learning rate scheduler. The training is terminated when the reconstruction loss on the validation data does not diminish for 50 epochs and the lowest validation loss is used for the final model.

\subsection*{Hyperparameter selection and model design optimization}
Model hyperparameters and architecture are optimized for laboratory experiment p4677 data using a Bayesian optimizer \cite<scikit-optimize package;>{head_tim_2021_5565057} to train thousands of models with slightly different designs. 
Optimization is performed for the two CED models for embedding the input AE and output frictional coefficient, respectively, and the Transformer model for predicting future output embeddings. 
For the output CED model, the number of convolutional layers $n^l_{out}$, filter size $f_{out}$, kernel size $k_{out}$, the number of codes in the VQ-VAE's codebook $n^{vq}_{out}$, the activation function $act_{out}$ and the L2 regularization weight $l2_{out}$ are optimized. 
Similarly for the input CED model, $n^l_{in}$, $f_{in}$, $k_{in}$, $n^{vq}_{in}$, $act_{in}$, and $l2_{in}$ are optimized. 
The hyperparameters for the Transformer include the number of encoder layers $n^{trsf}_{enc}$, the number of decoder layers $n^{trsf}_{dec}$, the number of attention heads $n^{h}_{attn}$, and the dropout rate $r_{drop}$  (see Table S1). 
The search space is comprised of a range of values for each variable listed. 
The objective function for hyperparameter minimization is the MSE between the target future friction coefficients and the predicted frictions on the validation data of experiment No. p4677 using AE history of 2.56 s to predict 0.256 s into the future. 
Each optimization attempt is run for 500 iterations, with the first 100 iterations using random parameter selections. 
The procedure is distributed onto 16 GPUs with 5 processes running per GPU (80 models trained simultaneously) to rapidly test the search space. The procedure is iterated to adjust the search spaces until the final values are not at the boundary limits. 
The optimized hyperparameters for the final model design are provided in  Supporting Information Table S1.

\subsection*{Laboratory experiment} 
We use laboratory data obtained from a bi-axial shear device. The device is a double-direct shear apparatus comprised of three blocks with two fault gouge layers, e.g., \cite{Marone98,Johnson2013acoustic,Rouet-Leduc_GRL_2017,JohnsonPA_PNAS_2021}.  The three block system is held in place by a fixed horizontal load, while the center block is driven by a piston oriented perpendicular to the horizontal load. The shear stress and friction on the gouge layers, from the center drive block and measured with a load cell, is an important experimental output. The simulated fault gouge comprises class IV spheres (dimension from 105–149 $\mu$ m). The initial layer thickness is 2 $\times$ 4 mm (two layers), and the roughened interfaces with the drive block have dimensions 10 cm $\times$ 10 cm. The drive block vertical displacement rate is 5 $\mu$ m/s, corresponding to a strain rate of approximately 1.2 $\times$ $10^{-3}$ /s. The apparatus is servo-controlled so that constant normal stress and displacement rate of the drive block are maintained at $\pm$0.1 kN and $\pm$0.1 $\mu$ m/s, respectively. The apparatus is monitored via computer to record load on the drive block and drive block displacement at 1 kHz. Once the system has been sheared to the point where it is in steady state conditions, the laboratory fault gouge layers fail in quasi-periodic cycles of stick and slip.

\subsection*{Laboratory Data Analysis} 
For the present analysis, experiment No. p4677 is used to train and validate the model, in which a fixed normal stress on the blocks of 2.5 MPa is maintained. Continuous AE broadcast from the fault zone were recorded via a Verasonics multichannel recording system and used as the input signal to the deep learning model, as described above. 
The coefficient of friction ($\mu$) was used as the target output signal. 
The entire experiment p4677 signals are split into 60\% training and 40\% validation data. 
These segments of signals are further split into sliding windows with the window size for inputs $l_{in}$, the window size for outputs $l_{out}$, the step size for the input windows $l_{step}$, and the lag between the start of the input and output windows $l_{lag}$. 
For the first step of training the CED models with instantaneous prediction data, $l_{in} = l_{out} = l_{step}$, $l_{lag} = 0s$, and $l_{in}$ is equal to the unit window length (0.256 s in the Results) of signals embedded into one latent vector. 
For the second step of training the Transformer in the latent space with future prediction data while keeping the pre-trained CED fixed, $l_{in}$ and $l_{out}$ are no longer required to be the same, $l_{lag} = l_{in}$ and $l_{step} = l_{out}$ to ensure zero overlapping between input and output windows. 
The values of 0.256 s, 0.512 s, 1.024 s, 1.536 s, 2.048 s, 2.56 s are tested for $l_{in}$ and $l_{out}$ in order to study the window size effect on the prediction accuracy of the future fault slips. 
The size of the datasets depends on the above length hyperparameters. 
For example, for future prediction data with $l_{in} = 2.56s$, $l_{out} = 0.256s$, $l_{lag} = 2.56s$ and $l_{step} = 0.256s$, there are 694 pairs of input and output signal windows in the training dataset and 459 pairs in the validation dataset. 
Experiment No. p4581 is used as the testing dataset, in which progressively larger normal loads were applied and the steady states are at 3, 4, 5, 6, 7, 8 Mpa normal stresses. In the following we focus on the 3MPa load analysis.

All input and output signals are normalized by subtracting the mean and dividing by the standard deviation using the statistics extracted from the training signal data. For the p4677 data, the statistics from the training signals are 8.878$\pm$21.018 for the input AE signals and 0.652$\pm$0.0413 for the output $\mu$. 
When making predictions using experiment p4581 data with increasing normal loads, the statistics are extracted from the first 40\% of the signals at each normal stresses for AE and 0.435$\pm$0.024 for $\mu$.

\section*{Results and Discussion}
We first compared the future prediction accuracy between the classification approach and the regression approach, as well as between using the full waveform and AE magnitude signals as model inputs. We found classification outperforms regression and therefore will focus on these results. The optimal model we found was using the classification approach with the AE magnitudes as inputs after hyperparameter optimization (Table S1). 

\begin{figure}[htbp]
\centering
\includegraphics[width=1\linewidth]{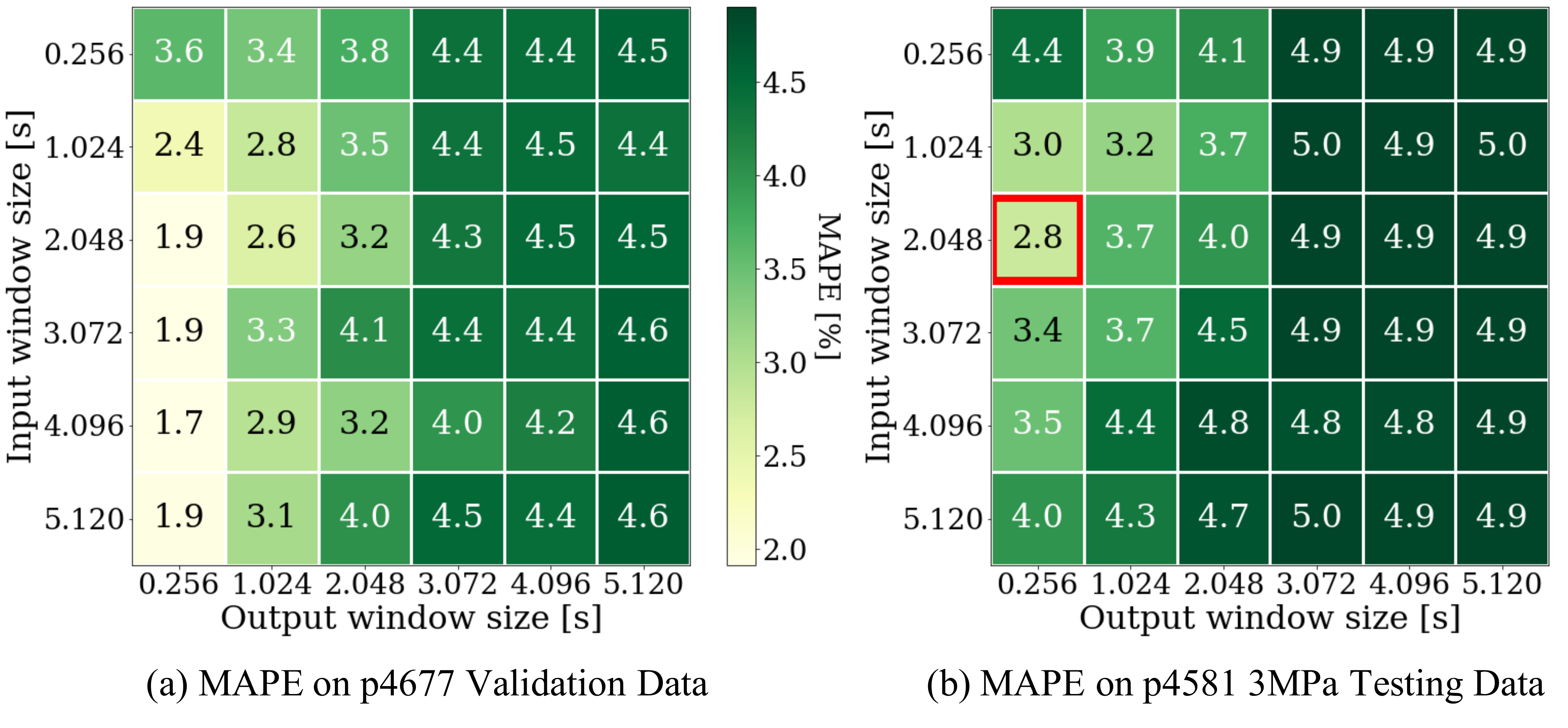}
\caption{Grid search for 36 different input and prediction window lengths, showing MAPE of future predictions using different input/output window sizes. The validation data (left) is from experiment p4677, 2.5 MPa applied load data with 60\%/40\% train/validation split, and the testing data (right) are from the experiment p4581 dataset conducted at 3 MPa. The window highlighted with the red rectangle has the smallest testing data error and these model predictions are plotted in Figure \ref{fig:ced_model_attn_1}. Supplementary matrices of MAPE results with finer window size (0.256 s step size) until 2.56 s is provided in Figure S1.}
\label{fig:ced_model_p4581_window_sizes}
\end{figure}

The future prediction accuracy using progressively increasing input and output window sizes on the validation and testing signals are tabulated in the matrix shown in Figure \ref{fig:ced_model_p4581_window_sizes}. 
A larger matrix showing more window sizes is provided in Figure S1 of Supporting Information. 
The Mean Absolute Percent Error (MAPE) values are noted in the matrix figures for validation and testing, respectively. We note that average MAPE/MSE scores are not a perfect indicator of model performance; however, they do provide an indication of prediction accuracy (and are commonly applied metrics).
The validation data set uses a portion of experiment p4677 data the model had not seen during training. 
The trained model is then applied to the p4581 testing data set for an experiment that were conducted at a slightly different load level (3 MPa vs. 2.5 MPa).
The results show for the validation set that near-term predictions are about the same with the exception of the shortest window length. In contrast, the testing data shows degrading prediction as the input window length increased beyond 2.048 s. Similarly, predictions further into the future degrade with window length.
The best results for the training data are in the range of 1.024 - 2.048 seconds input and 0.256 s prediction window length. Predictions degrade with longer window lengths into the future.  In summary, the model predictions are best for short future time sequences, and predictions beyond approximately 2.5 s are less good.

\begin{figure}[htbp]
\centering
\includegraphics[width=0.99\linewidth]{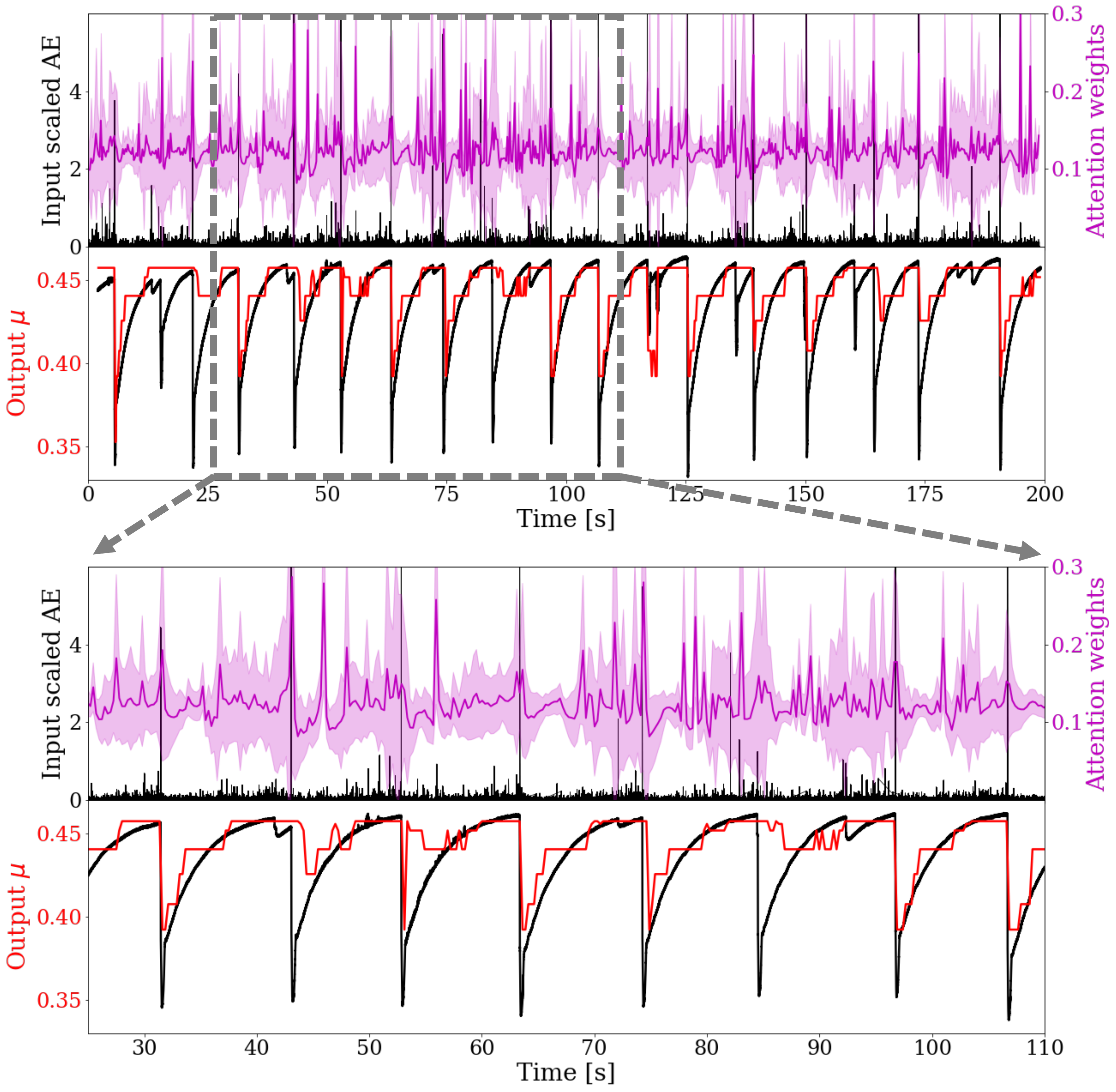}
\caption{Attention scores shown with the AE and predicted friction values for the model shown in the red rectangle in Figure \ref{fig:ced_model_p4581_window_sizes}. In each figure the upper panel shows the mean attention score  $\pm$ 1 standard deviation when predicting segments of 0.256 s window using 2.048 s history. The lower panel is the corresponding friction prediction. The bottom figure shows an expanded view from 25s to 110s to better highlight the time series details. In the 8 significant fault failures during this period, the timing of 5 failures are predicted at the correct time step, 2 are predicted after the slip event, and 1 is missed by the model.}
\label{fig:ced_model_attn_1}
\end{figure}

\begin{figure}[htbp]
\centering
\includegraphics[width=0.95\linewidth]{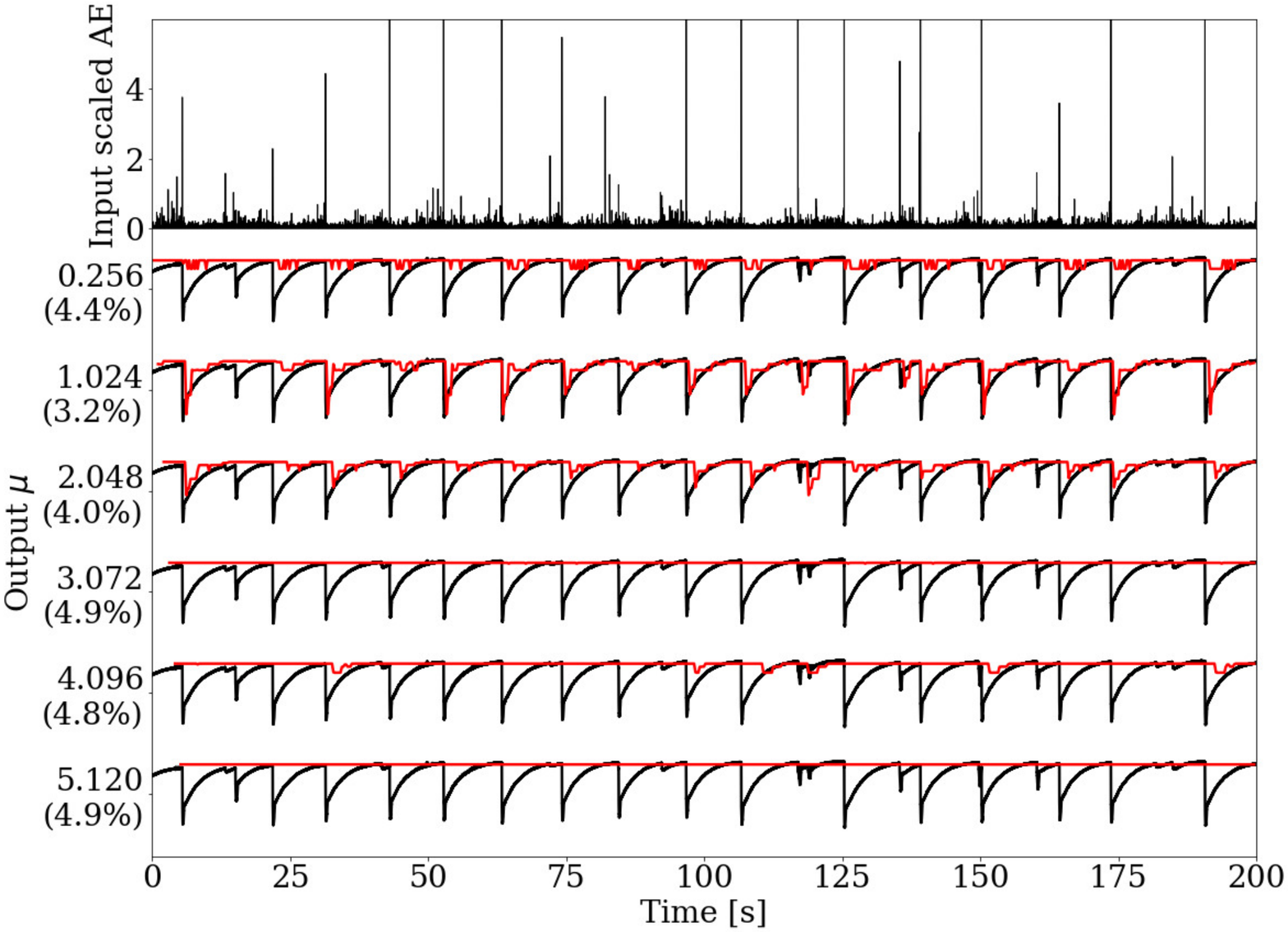}
\caption{Test data (p4581) and the model predictions of the friction using equal length input and output window sizes (the diagonal of Figure \ref{fig:ced_model_p4581_window_sizes}b). The friction prediction is shown in red and the experiment measurements in black. The y-axis labels shows the input and output window lengths 0.256s, 1.024s, 2.048s, 3.072s, 4.096s, 5.12s and the MAPE error in  parentheses. More prediction examples are provided in Supplementary materials Figure S2 and S3.
}
\label{fig:ced_model_p4581_test_equal}
\end{figure}

Figure \ref{fig:ced_model_attn_1} shows the predictions for the time window exhibiting the best MAPE---the matrix element denoted with the red box in Figure \ref{fig:ced_model_p4581_window_sizes}.  Figure \ref{fig:ced_model_p4581_test_equal} shows predictions for the diagonal elements shown in the right panel in Figure \ref{fig:ced_model_p4581_window_sizes}. In Figure \ref{fig:ced_model_attn_1} one can see that the predictions are reasonably good overall. Many frictional failures are predicted, as are the succeeding recoveries. For some slip events, the event is not well predicted and the onset of recovery is therefore poorly predicted, e.g., events at approximately 44 and 54 sec. Large precursors may be predicted (e.g., approximately 120 sec) or missed entirely (e.g., 15 sec). While a slip event may be predicted, the full friction failure magnitude is not as was observed in previous works applying many different models, e.g.,\cite{JohnsonPA_PNAS_2021}.  In general the predictions improve through the slip cycle. The fact that the model is able to predict the failure event just before it takes place is encouraging. The onset of frictional recovery post-failure is also predicted in many cases suggesting there is knowledge contained in the AE regarding the onset of the future labquake cycle.  The intense acoustic emission associated with the failure event apparently does not entirely erase fault system memory.

From the self-attention calculated in the Transformer latent-space, the relative importance of different parts of the AE input for future predictions can be analyzed by plotting the attention scores to visualize their distribution with respect to the AE history. The top panel of Figure \ref{fig:ced_model_attn_1} shows the mean attention scores and their standard deviation, computed from successive sliding input windows. The attention scores represent the average of the attention heads for a given input window. An expanded  view of eight stress cycles is shown in the lower panel to better visualize the variance in the signal and the accuracy of the friction predictions. Where the attention scores are largest indicates time intervals when the AE signal is most important for the future friction predictions. In general, the self attention scores indicate the model predictions are better in time intervals where impulsive precursors occur immediately preceding failure but also where the scores fluctuate the most. 

The future predictions using the p4581 testing data for successively larger input and output windows as shown along the diagonal in the right panel in Figure \ref{fig:ced_model_p4581_window_sizes} are presented in Figure \ref{fig:ced_model_p4581_test_equal}. The time series data illustrate what the MAPE values listed in  Figure \ref{fig:ced_model_p4581_window_sizes} indicate. The 1.024 s input and output underscore the best prediction capability, and the prediction degrades progressively as the prediction time window size increases.  Looking at the 0.256s results the predictions are quite poor, possibly due to such a small input window size. For the time windows longer than 3 seconds there appears to be insufficient information contained in the AE signal to predict anything, but this also may reflect the model generalization. We note that poorer accuracy for longer window lengths could arise from the limitations of the Transformer model to capture long-term dependency in the latent space (embedding) of signals, although proven to be very successful in applications for natural language processing. 

Previous work shows good instantaneous prediction capability contained in AE continuous waveforms. Signal energy was identified as key to instantaneous and time-to-failure prediction using both decision tree and deep learning models in the laboratory experiment data \cite<e.g.,>{Rouet-Leduc_GRL_2017, JohnsonPA_PNAS_2021} and for slow slip in Earth \cite<e.g.,>{Hulbert_NatGeo_2019, Johnson_GRL_2021}.  The analysis described here shows there is also information contained in the continuous signal regarding the immediate future behavior. The model attention scores focus on the region preceding frictional failure where precursors occur. This is logical near failure, based on our knowledge of classical seismic precursors in the laboratory \cite<e.g.,>{Johnson2013} and those observed often, but not always, in Earth \cite{scholz,Bouchon2013}. The results also indicate that within the inter-event portion of the earthquake cycle, near term friction prediction is also possible. Perhaps most interesting is that the seismic signal contains information predictive of the fault friction immediately beyond the failure event when one might expect the system to be unstable. The attention scores do not shed light as to why this is the case, but it is truly intriguing and future endeavors might provide clues as to what features of the signal contain this information. Applying  machine learning approaches such as that described in this work, to continuous seismic signals as well as other geophysical data, will continue advancing our understanding of fault activity while advancing earthquake hazards assessment.

%

%
%
%
%
%
%

\section*{Acknowledgements}
KW and PAJ acknowledge support by the U.S. Department of Energy, Office of Science, Office of Basic Energy Sciences, Chemical Sciences, Geosciences, and Biosciences Division under grant 89233218CNA000001. KW also acknowledges support by the Center for Nonlinear Studies (CNLS) at Los Alamos National Laboratory. CWJ and KCB acknowledge Institutional Support (Laboratory Directed Research and Development) at Los Alamos National Laboratory. The authors declare no competing interests. We thank Chris Marone for the laboratory data.


%
%

\bibliography{biblio.bib}

%
%
%
%
%

\end{document}


\maketitle


\noindent\textbf{Contents of this file}
\begin{enumerate}
\item Figures S1 to S4.
\item Table S1.
\end{enumerate}

\begin{table}
\centering
\begin{tabular}{ |l|c|l|c| }
\hline
Hyperparameter & Symbol & Choices & Value \\ \cline{1-4}
Output number of convolutional filters & $f_{out}$ & 4, 8, 16, 32, 64 & 8 \\ \cline{1-4}
Output number of encoder-decoder layers & $n^l_{out}$ & 3, 4, 5 & 4 \\ \cline{1-4}
Output kernel size & $k_{out}$ & 3, 5, 7, 9, 11 & 7 \\ \cline{1-4}
Output number of vector-quantized codes & $n^{vq}_{out}$ & 16, 32, 48, 64, 96, 128 & 16 \\ \cline{1-4}
Output activation function & $act_{out}$ & ReLU, Leaky ReLU, ELU, GELU & GELU \\ \cline{1-4}
Output L2 penalty coefficient & $l2_{out}$ & 1E-6, 1E-5, 1E-4, 1E-3, 1E-2 & 1E-4 \\ \cline{1-4}
Input number of convolutional filters & $f_{in}$ & 8, 16, 32, 64 & 32 \\ \cline{1-4}
Input number of encoder-decoder layers & $n^{l}_{in}$ & 3, 4, 5 & 4 \\ \cline{1-4}
Input kernel size & $k_{in}$ & 3, 5, 7, 9, 11 & 9 \\ \cline{1-4}
Input number of vector-quantized codes & $n^{vq}_{in}$ & 16, 32, 48, 64, 96, 128 & 96 \\ \cline{1-4}
Input activation function & $act_{in}$ & ReLU, Leaky ReLU, ELU, GELU & GELU \\ \cline{1-4}
Input L2 penalty coefficient & $l2_{in}$ & 1E-6, 1E-5, 1E-4, 1E-3, 1E-2 & 1E-4 \\ \cline{1-4}
Transformer number of encoder layers & $n^{l}_{enc}$ & 1, 2, 3, 4 & 2 \\ \cline{1-4}
Transformer number of decoder layers & $n^{l}_{dec}$ & 1, 2, 3, 4 & 2 \\ \cline{1-4}
Transformer number of attention heads & $n^{h}_{attn}$ & 1, 2, 4, 8 & 4 \\ \cline{1-4}
Transformer dropout rate & $r_{drop}$ & 0.1, 0.2, 0.3, 0.4 & 0.1 \\ \cline{1-4}
\end{tabular}
\caption{Hyperparameter optimization involved in the deep learning model. The table includes the descriptions of the hyperparameters, the search space and the final adopted values after optimization.}
\end{table}

\begin{figure}[htbp]
\centering
\includegraphics[width=1.0\linewidth]{Figures/future_predict_small_window_sizes_mape.pdf}
\caption{MAPE of future predictions using different input/output window sizes, the training data is the p4677 2.5MPa data with 60\%/40\% train/valid split, and the testing data are from a different experiment, the p4581 3MPa dataset.}
\label{label_here}
\end{figure}

\begin{figure}[htbp]
\centering
\includegraphics[width=1.0\linewidth]{Figures/window_prediction_in2048_p4581_test.png}
\caption{
Test data (p4581) and the model predictions of the friction using 2.048s input and increasing output window sizes (the third row of Figure S1 b). The friction prediction is shown in red and the experiment measurements in black. The y-axis labels shows the output window lengths 0.256s, 0.512s, 1.024s, 1.536s, 2.048s, 2.560s and the MAPE error in  parentheses.}
\label{label_here}
\end{figure}

\begin{figure}[htbp]
\centering
\includegraphics[width=1.0\linewidth]{Figures/window_prediction_equal_p4581_test_2560.png}
\caption{Test data (p4581) and the model predictions of the friction using equal length input and output window sizes (the diagonal of Figure S1 b). The friction prediction is shown in red and the experiment measurements in black. The y-axis labels shows the input and output window lengths 0.256s, 0.512s, 1.024s, 1.536s, 2.048s, 2.560s and the MAPE error in  parentheses.}
\label{label_here}
\end{figure}
